\documentclass[%
 reprint,
 amsmath,amssymb,
 aps,
pra,
]{revtex4-2}

\usepackage[normalem]{ulem}  % add in preamble

\usepackage{graphicx}% Include figure files
\usepackage{dcolumn}% Align table columns on decimal point
\usepackage{appendix}
\usepackage{bm}% bold math
\usepackage{siunitx}
\usepackage{xcolor}
\usepackage{CJK}
\usepackage{hyperref}% add hypertext capabilities
\usepackage{physics}

\newcommand{\subref}[2]{\hyperref[#1]{\ref*{#1}(#2)}}
\definecolor{refblue}{HTML}{2E2E91}
\hypersetup{colorlinks=true, citecolor=refblue, urlcolor=refblue, linkcolor=refblue}

\begin{document}

\preprint{APS/123-QED}

\title{A High-Power Clock Laser Spectrally Tailored for High-Fidelity Quantum State Engineering}% Force line breaks with \\
% \thanks{}%

% \author{Lingfeng Yan, Stefan Lannig, William R. Milner, Max\CJKkern~N.~Frankel, Ben Lewis, Dahyeon Lee, Kyungtae Kim, and Jun Ye}
%  % \altaffiliation[Also at ]{CU}%Lines break automatically or can be forced with \\
% % \author{Lingfeng Yan}%
%  \email{lingfeng.yan@colorado.edu, ye@jila.colorado.edu}
\author{Lingfeng Yan} 
\email{lingfeng.yan@colorado.edu}
\author{Stefan Lannig}
\author{William R. Milner}
\author{Max~N.~Frankel}
\author{Ben Lewis}
\author{Dahyeon Lee}
\author{Kyungtae Kim}
\author{Jun Ye}
% \author{Stefan Lannig, William R. Milner, Max\CJKkern~N.~Frankel, Ben Lewis, Dahyeon Lee, Kyungtae Kim, and Jun Ye}
\email{ye@jila.colorado.edu}
\affiliation{%
 JILA, National Institute of Standards and Technology and University of Colorado and Department of Physics, University of Colorado, Boulder, CO 80309, USA %\textbackslash\textbackslash
}

\date{\today}% It is always \today, today,
             %  but any date may be explicitly specified

\begin{abstract}
Highly frequency-stable lasers are a ubiquitous tool for optical frequency metrology, precision interferometry, and quantum information science. While making a universally applicable laser is unrealistic, spectral noise can be tailored for specific applications. \textcolor{black}{Here we report a high-power 698 nm clock laser with a maximum output of \SI{4}{W} and minimized frequency noise up to a few kHz Fourier frequency, together with long-term instability of $3.5 \times 10^{-17}$ at one to thousands of seconds.} The laser frequency noise is precisely characterized with atom-based spectral analysis that employs a pulse sequence designed to suppress sensitivity to intensity noise. This method provides universally applicable tunability of the spectral response and analysis of quantum sensors over a wide frequency range. With the optimized laser system characterized by this technique, we achieve an average single-qubit Clifford gate fidelity of up to $F_1^2 = 0.99964(3)$ when simultaneously driving 3000 optical qubits with a homogeneous Rabi frequency ranging from \SI{10}{Hz} to $\sim$$\SI{1}{kHz}$. This result represents the highest single optical-qubit gate fidelity for large number of atoms.

\end{abstract}

%\keywords{Suggested keywords}%Use showkeys class option if keyword
                              %display desired
\maketitle

%\tableofcontents
\section{\label{sec:Introduction}Introduction}
Preparing and manipulating large-scale quantum systems with high fidelity is critical for many applications using cold atoms and molecules. To cool, trap, and ultimately control quantum states of these particles, lasers are employed. Improving the spectral properties of lasers, thus advancing the fidelity and robustness of control for these quantum systems, will open the door for new physics studies~\cite{YeZoller}. 

% Trapped ion platforms have achieved the highest single- and two-qubit gate fidelities, see for example Refs.~\cite{Be2016PRL, Clark2021PRL, Leu2023PRL, AQT_website, quantinuum2024demonstration,loschnauer2024scalable}. Neutral atom-based qubit arrays, which are more easily scalable and are approaching competitive fidelities~\cite{evered2023high, ma2023high, bluvstein2024logical,radnaev2024universal,tsai2024benchmarking, muniz2024high}, will benefit from lasers engineered with improved frequency noise. For optical qubits based on high-quality-factor optical transitions of alkaline-earth atoms, which have emerged as excellent platforms for both quantum metrology and information~\cite{eckner2023realizing,finkelstein2024universal,cao2024multi}, the gate fidelity is directly affected by the spectral properties of an optical local oscillator that provides the driving field. This has in part limited optical single-qubit fidelities to $< 0.999$~\cite{Lis2023PRX_omg, madjarov2020high}.
\textcolor{black}{Optical qubits based on high-quality-factor optical transitions have emerged as excellent platforms for both quantum metrology and information. Trapped ion platforms have achieved the highest single- and two-qubit gate fidelities, see for example Refs.~\cite{Be2016PRL, Clark2021PRL, Leu2023PRL, AQT_website, quantinuum2024demonstration,loschnauer2024scalable}. Neutral atom-based qubit arrays, which are more easily scalable, are approaching competitive fidelities~\cite{evered2023high, ma2023high, bluvstein2024logical,radnaev2024universal,tsai2024benchmarking, muniz2024high,eckner2023realizing,finkelstein2024universal,cao2024multi,chiu2025continuous}. Importantly, these atomic systems feature excellent coherence times, for example, exceeding \SI{100}{s} for clock transition in Sr system~\cite{kim2025atomic, ma2025enhancing}, thus allowing a deep circuit even with a relatively ``slow'' Rabi frequency of a few kHz. The gate fidelities of optical qubits are directly affected by the spectral properties of an optical local oscillator that provides the driving field. This has in part limited optical single-qubit fidelities to $< 0.999$~\cite{Lis2023PRX_omg, madjarov2020high}. }
Even for alkali atoms employing hyperfine qubit states with energy separation in the radio frequency domain, spectral filtering of the Rydberg lasers used for entangling operations was important for achieving two-qubit gate fidelities $> 0.95$~\cite{levine2018highfidelity}. Continuing to improve the fidelity of gate operations will reduce the number of physical qubits required and the operation time to correct errors~\cite{oskin2002practical,knill2010quantum}, paving the way for the implementation of useful quantum error correction~\cite{preskill1998reliable,nielsen2010quantum}. 

Beyond quantum information, improved lasers expand the scope of quantum metrology and many-body physics studies. For optical clocks, an improved laser coherence time directly translates into enhanced clock precision, allowing for more stringent bounds on fundamental physics tests~\cite{kennedy2020precision, kolkowitz2016gravitational} and a finer resolution for studies of novel clock systematics~\cite{hutson2024observation}. Using high-fidelity entangling operations to introduce quantum correlations, clock stability can be further enhanced beyond the standard quantum limit~\cite{Vuletic2020,eckner2023realizing, robinson2024direct}. Quantum gases of molecules also benefit from improved lasers, where stimulated Raman adiabatic state transfer to the rovibrational ground state is prone to the limitation imposed by optical phase noise~\cite{Ni2008,maddox2024enhanced}. 
\textcolor{black}{Improved laser pulse fidelities will also unlock new opportunities to probe rich light-matter interactions (such as collective Lamb shift and superradiance/subradiance phenomena~\cite{hutson2024observation, Masson2024superradiance}), a general XYZ spin model to be used to produce spin squeezing~\cite{Mamaev2024superexchange, milner2024coherent, miller2024KRbFluoquet, lee2024easy_plane}, a new interface between general relativity and quantum dynamics~\cite{Chu2025Gravity}. Extending the scope, macroscopic ensemble-based fundamental physics experiments, such as optical transition-based atom interferometry~\cite{Graham2013LMT,Chiarotti2022LMT_limit,baynham2025prototype}, would directly benefit from improved pulse fidelity. These examples illustrate the broader scientific potential of precise and coherent optical control enabled by improved laser systems.}

Realizing high-fidelity quantum state engineering requires precise control of field-matter interactions. 
For the driving field, a sufficiently large amplitude for fast quantum state rotations (gates) is typically desired to outpace decoherence effects and reduce sensitivity to laser noise. Of course, simply performing fast rotations is not enough to ensure high fidelity; the amplitude and phase noise of the driving field can significantly limit performance~\cite{geva1995relaxation, makhlin2003dephasing,ithier2005superconductingQdecoherence, chen2012general, yan2013rotating,1/fsolidQI,yoshihara2014fluxQNoiseSpectroscopy,1/fsolidQI,jing2014decoherence,masterclockstabilityqubit2012npj,green2012high,green2013arbitrary,2018AnalysisRydbergImperfection,day2022limits,jiang2023sensitivity, tsai2024benchmarking}. 
For drive pulses resonant with qubits, phase noise near the Rabi frequency is especially harmful to fidelity. As the Rabi frequency increases, which is often needed for versatile manipulations of quantum states such as Floquet engineering~\cite{choi2020robust}, it becomes necessary to suppress the drive-field noise at higher Fourier frequencies relative to the carrier. Additionally, attention to low-frequency phase noise is crucial, as it constrains the long-term stability of the local oscillator and, ultimately, the maximum coherence time between the drive field and the qubits. 

Achieving superior frequency stability over a large Fourier spectral range typically requires stabilizing a laser to a high-finesse reference cavity. Cryogenic silicon cavities represent the state of the art in stable cavity technology and a long phase-coherence time of tens of seconds has been demonstrated~\cite{matei20171}. However, even such high-performance optical cavities can have limitations, for example arising from the limited use of optical power to achieve long-term stability at the cost of relatively higher shot noise contributions~\cite{Robinson:19}. Furthermore, a high-performance cavity is usually optimized for a relatively specific wavelength range. To transfer this stability to lasers at other wavelengths, an optical frequency comb is necessary~\cite{cundiff2003colloquiumfreqComb, ludlow2015opticalClock}. This transfer process involves phase-locking a tooth of the frequency comb to the cavity-stabilized continuous-wave laser. The high-frequency phase noise of the stabilized comb is generally constrained by a white phase noise floor, primarily due to the shot noise resulting from the limited optical power available for the individual frequency comb tooth among many neighboring lines~\cite{Newbury:07}.

To mitigate high-frequency noise in lasers, various stabilization techniques have been developed. Typically, lasers are directly locked to a reference cavity using a fast feedback loop, which reduces noise within the loop's bandwidth. The bandwidth of the feedback loop is ultimately limited by the overall time delay between sensing and actuation, and an optimized bandwidth of a few MHz can be achieved for high-performance feedback systems~\cite{hall1984external, Zhu:93, Schoof:01, LEGOUET2009977,Appel_2009, Gatti:15, endo2018highbandwidthPDHfeedback, preuschoff2022wideband}. In case the servo bandwidth is limited, reduction of high-frequency noise can be achieved with two effective approaches: using the transmission from a narrow-linewidth cavity or using a feedforward technique to cancel noise within a certain frequency band. The former approach relies on the narrow-linewidth cavity to filter out phase noise of the transmitted light beyond the cavity bandwidth~\cite{Hald2005Optica_noisefiltering, Nazarova08_low_noise_laser}. An improvement of gate fidelity using this technique has been demonstrated~\cite{levine2018highfidelity}. However, the power of light transmitted through such cavities is often limited to avoid thermal instability and optical damage to the coatings due to intense intracavity fields. 

The feedforward technique offers an alternative approach to reducing high-frequency noise by acting directly on detected frequency noise without going through a feedback loop which may introduce instability at high bandwidth~\cite{Bagheri:09_feedforward,Aflatouni2010feedforward,aflatouni2012feedforward,lintz2017note_feedforward,li2022activefeedforward, chao2024PDHfeedforward,denecker2024measurement}. 
\textcolor{black}{When electronic delay is limiting the speed of a servo loop, feedforward control, supplemented by an optical delay line that matches the electronic delay, can effectively extend the noise suppression bandwidth. However, feedforward systems are often more susceptible to unforeseen disturbances than traditional feedback systems.}
The decision on laser noise reduction should be made based on the practical requirements of a quantum system one wishes to control.

In this article, we report the development of an ultra-stable, high-power laser of multiple Watts that achieves the stability of our state-of-the-art clock laser~\cite{oelker2019demonstration}. This allows us to demonstrate a high-fidelity quantum state preparation equivalent to an average single-qubit gate fidelity of up to 0.99964(3) at a Rabi frequency of 0.74 kHz for 3000 atoms. Two stable optical reference cavities, a cryogenic silicon cavity at \SI{1.5}{\micro m} and a room-temperature ULE cavity at \SI{0.7}{\micro m}, work in tandem to achieve superior performance in reducing laser noise at low and high Fourier frequency, respectively. A phase-stabilized optical frequency comb provides a low-noise stability transfer to bridge the spectral gap between the two cavities. An additional phase-locked loop then transfers the stability of this spectrally tailored laser to a high-power local oscillator for fast quantum state manipulations of atoms. We precisely characterize the laser frequency noise using atom-based spectral analysis to report outstanding fidelity for randomized benchmarking Clifford gates operating at Rabi frequencies ranging from tens of Hz to $\sim\SI{1}{kHz}$. Notably, we achieve this high fidelity nearly uniformly for 3000 atoms confined in the 3D lattice, confirmed with imaging spectroscopy~\cite{marti2018imaging,Milner2023PRA_imaging}. 

We proceed in Sec.~\ref{sec: experiment} with a description of the experimental platform for the quantum system under interest. In Sec.~\ref{sec:laser frequency noise reduction} we present a detailed scheme for laser frequency noise stabilization and the corresponding noise model of the spectrally tailored laser. In Sec.~\ref{sec: on-site frequency noise measurement pulses} we develop a pulse sequence with reduced sensitivity to laser intensity noise but a localized sensitivity to a narrow frequency noise band such that atoms can be turned into a quantum optical spectrum analyzer to perform on-site laser frequency noise measurements. We experimentally verify its performance and use it to validate the laser noise model introduced in Sec.~\ref{sec:laser frequency noise reduction}.
The reduced high-frequency noise of our clock laser system enables high-fidelity single-qubit operations which we quantitatively characterize via randomized benchmarking in Sec.~\ref{sec: RB}.

\section{Laser-atom experimental platform\label{sec: experiment}}
\begin{figure}[hbtp!]
    \centering
    \includegraphics[width=\columnwidth]{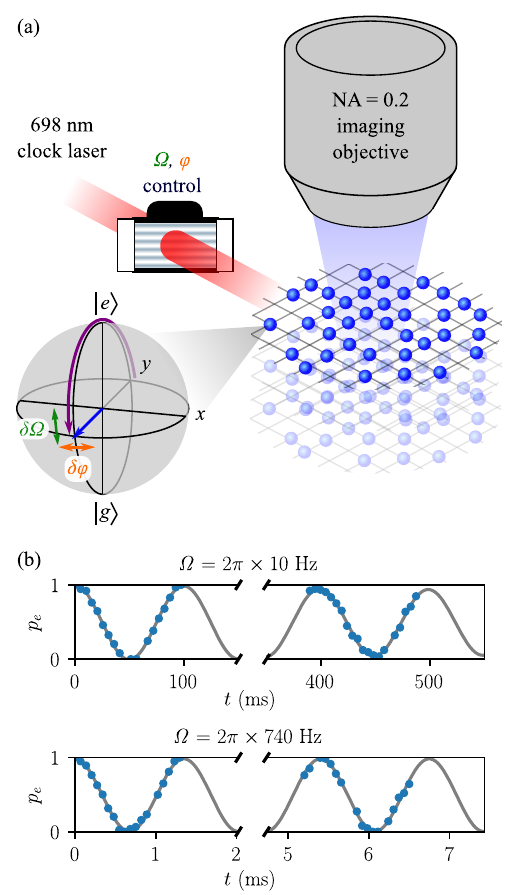}
    \caption{
	\textbf{3D optical lattice clock platform for high-fidelity quantum state engineering.}
    (a) Illustration of the experimental setup: A \SI{698}{nm} clock laser (red beam) drives the clock transition $|g = {}^1S_0\rangle \leftrightarrow |e = {}^3P_0 \rangle$ in ${}^{87}\text{Sr}$ atoms (blue spheres) that are confined in the ground band of a 3D \SI{813}{nm} magic wavelength optical lattice (gray lines). 
    Vertical absorption imaging provides access to spatial information at a resolution of $\SI{1.3}{\micro m}$. During clock pulses, the trajectory (purple line) of the atomic state (blue arrow) on the Bloch sphere is controlled by the laser phase $\varphi$ and Rabi frequency $\Omega$, both modulated by an acousto-optic modulator (AOM). 
    Intensity fluctuations induce variations $\delta\Omega$ in the Rabi frequency, affecting the pulse area, while laser phase noise $\delta\varphi$ causes errors in the orientation of the rotation axis. 
    \textcolor{black}{(b) Rabi flopping at $\Omega=2\pi \times \SI{10.031(6)}{Hz}$ (top) and $\Omega=2\pi \times \SI{741.0(6)}{Hz}$ (bottom) illustrate the range of Rabi frequencies achievable on the clock transition in our setup. Blue points are measurements of the excitation fraction $p_e$ extracted from single experiments and the gray line is a sinusoidal fit including an exponential decay.} 
    }
    \label{fig:fig1}
\end{figure}

State-prepared atoms confined in an optical lattice represent a well-controlled quantum system that can be used to characterize laser frequency noise over a broad spectral range and demonstrate high-fidelity quantum state engineering. Figure~\ref{fig:fig1}(a) presents a schematic of our experimental setup, featuring a three-dimensional optical lattice clock~\cite{campbell2017fermi}. A spin-polarized, deeply degenerate Fermi gas of $^{87}\text{Sr}$ atoms~\cite{sonderhouse2020thermodynamics} is loaded into the ground band of a retro-reflected \SI{813}{nm} magic wavelength optical lattice~\cite{ye2009magicwavelength}, which quantizes the motional states of atoms along all three orthogonal directions.
In this work, we operate at a lattice depth of $(V_x, V_y, V_z) \approx (45, 44, 41)E_r$, where $E_r \approx h \times\SI{3.5}{kHz}$ denotes the recoil energy for \SI{813}{nm} photons and $h$ the Planck constant. At this depth, atoms are confined within separate lattice sites, effectively isolated as two-level systems with negligible tunneling and superexchange interactions~\cite{milner2024coherent}. Under this operating condition, the atomic coherence time is \SI{7}{s}, limited primarily by Raman scattering of lattice photons~\cite{Hutson2019PRLshallow}.

The clock states of the atoms are manipulated by an intensity- and phase-controlled \SI{698}{nm} laser, which propagates horizontally to drive the ``clock" transition between the ground state $\ket{5s^2~^1\text{S}_0~m_F=-9/2}~(\ket{g})$ and the metastable excited state $\ket{5s5p~^3\text{P}_0~m_F=-9/2}~(\ket{e})$ globally. To achieve a uniform Rabi frequency across the entire atomic ensemble of $\sim 3000$ atoms within a spatial region of $\sim10\times10\times\SI{10}{\micro m^3}$, we use a sufficiently large beam \textcolor{black}{1/e$^2$} radius of \SI{770}{\micro\meter}. With this beam size, to scale up the clock Rabi frequency to about a few kHz we need to increase the optical power incident on the atoms to $>\SI{100}{mW}$. To compensate for beam propagation losses, it is desirable to reach an optical power $>\SI{1}{W}$ for the stabilized clock laser. \textcolor{black}{We note this power level is also sufficient to globally address thousands of atoms in a two-dimensional lattice or tweezer array if using a top-hat rather than a Gaussian beam.} \textcolor{black}{ However, we emphasize that for applications requiring large-scale transverse phase coherence, such as one-dimensional optical lattice clocks and atom interferometers, the wavefront quality of the top-hat beam can be corrected and verified.} High-intensity \textit{in-situ} absorption imaging, implemented along the vertical axis through an objective lens, allows spatial readout of atomic density distributions with a resolution of \SI{1.3}{\micro\meter}~\cite{marti2018imaging,Milner2023PRA_imaging}. With the setup described in this article we demonstrate driving Rabi oscillations with frequencies ranging from a few Hertz to approximately \SI{1}{kHz}, illustrated in Fig.~\ref{fig:fig1}(b) at Rabi frequencies $\Omega=2\pi \times \SI{10.031(6)}{Hz}$ and $\Omega=2\pi \times \SI{741.0(6)}{Hz}$.

\section{\label{sec:laser frequency noise reduction}Laser frequency noise reduction}
For this 3D lattice-based quantum platform, the achievable Rabi frequencies are within the range of a few kHz. A feedback loop of $\approx$\SI{1}{MHz} bandwidth is sufficiently fast to reduce laser noise to the cavity’s noise floor within the spectral region of interest. We thus utilize the Pound-Drever-Hall (PDH) locking technique~\cite{drever1983laser} to stabilize a \SI{698}{nm} external cavity diode laser (ECDL) to a narrow-linewidth ultra-low expansion (ULE) cavity as a reference with low high-frequency noise. For this laser to benefit from the long-term stability of a cryogenic silicon cavity, we employ a specifically designed stability transfer loop that utilizes a frequency comb to steer the \SI{698}{nm} laser to the silicon cavity at \SI{1.5}{\micro m}.
The resulting laser sent to the atoms is spectrally tailored such that it inherits the cryogenic silicon cavity's long-term stability and the ULE cavity's low noise at high frequency, making it suitable for high-fidelity quantum state engineering.

\begin{figure*}[hbtp!]
    \centering
    \includegraphics[width=\textwidth]{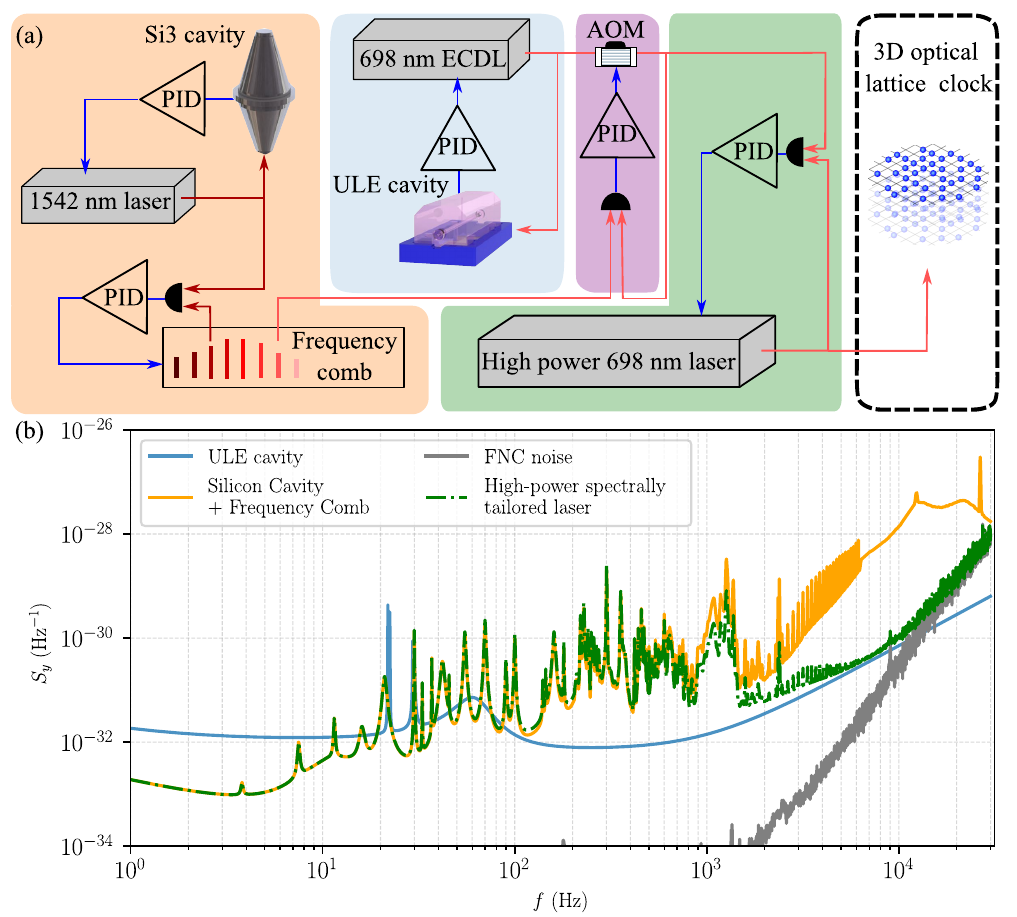}
    \caption{
	    \textbf{Clock laser frequency noise stabilization.}
        (a) A simplified schematic of the stabilization chain from the reference cavities to the atoms. Servo feedback is indicated by blue arrows and laser light is depicted as red arrows.
        A \SI{1542}{nm} laser is PDH locked to the cryogenic silicon cavity (``Si3 cavity") with a bandwidth of \SI{100}{kHz}, which provides the low-frequency stability ($f\lesssim\SI{100}{Hz}$) of the entire system. A self-referenced frequency comb is phase-locked to the Si3 cavity-stabilized \SI{1542}{nm} laser with a bandwidth of \SI{600}{kHz}. 
        High-frequency noise ($f\gtrsim\SI{1}{kHz}$) is reduced by PDH-locking a \SI{698}{nm} ECDL to a ULE cavity with \SI{1}{MHz} bandwidth (blue shaded region). The stability transfer at low Fourier frequency is achieved by phase-locking the ULE cavity-stabilized light to the Si3 cavity-stabilized comb via actuating on an AOM with a bandwidth of \SI{500}{Hz} (purple shaded region). This represents a low-power ($\approx$ \SI{10}{mW}) spectrally tailored laser with optimal noise performance. Finally, the light delivered to the atoms is provided by a high-power ($\gtrsim\SI{1}{W}$) laser phase-locked to the stabilized \SI{698}{nm} light with a bandwidth of \SI{400}{kHz} (green shaded region).
        (b) A budget of fractional laser frequency noise PSD ($S_y$) for different components of the laser system.
        For the spectrally tailored laser model shown here (green), the stability transfer bandwidth is chosen to be \SI{500}{Hz}, such that the low-frequency stability is inherited from the Si3 cavity-stabilized comb (orange) and the high-frequency stability from the ULE cavity-stabilized laser (blue). 
        At frequencies above $\sim\SI{10}{kHz}$, the in-loop noise from fiber noise cancellation (FNC) loops (gray) introduces additional noise.
        }
    \label{fig:fig2}
\end{figure*}

Figure~\ref{fig:fig2}(a) shows the block diagram of the stabilization scheme and Fig.~\ref{fig:fig2}(b) plots the fractional frequency noise power spectral density (PSD), $S_y$, relative to the carrier. \textcolor{black}{$S_y$ is related to the frequency noise power spectral density $S_\nu$ by $S_y=S_\nu/\nu_0^2$, where $\nu_0$ is the optical carrier frequency.} A $\SI{1542}{nm}$ laser is locked to a \SI{21}{cm} cryogenic silicon cavity (``Si3 cavity'') operating at $\SI{124}{K}$~\cite{matei20171}. The stability of the Si3 cavity has reached its thermal noise floor with an instability of 3.5$\times10^{-17}$ for 1 to over 1000 s interrogation time~\cite{oelker2019demonstration}. This cavity is the origin of the long-term stability for the clock laser and its stability is transferred to $\SI{698}{nm}$ via a self-referenced femtosecond Er-fiber frequency comb (orange shaded region in Fig.~\ref{fig:fig2}(a)). The noise model for the Si3 cavity-stabilized comb, based on a cross-correlation evaluation against two other independent lasers (see Appendix \ref{Append_A_1:Si}), is plotted as the orange solid line in Fig.~\ref{fig:fig2}(b).

To reduce the high-frequency noise of the final spectrally tailored laser, the \SI{698}{nm} ECDL is PDH-locked to a \SI{40}{cm} ULE cavity with a thermal noise floor of $1\times10^{-16}$~\cite{mattS2012highPrecisionClock, Nicholson2012Clk1e-17}, stabilized via a high-bandwidth feedback loop (blue shaded region in Fig.~\ref{fig:fig2}(a)). The \SI{1}{MHz} bandwidth PDH lock is capable of reducing the laser's in-loop frequency noise down to the detection noise floor within \textcolor{black}{\SI{50}{kHz}}, as detailed in Appendix \ref{Append:A2:ULE caivty}. The model of ULE cavity-stabilized laser is plotted as the blue line in Fig.~\ref{fig:fig2}(b).

To combine the long-term stability of the Si3 cavity and short-term stability of the ULE cavity, the ULE cavity-stabilized laser is phased-locked to the Si3 cavity-stabilized frequency comb at low frequencies by actuating on an AOM (shown in the purple shaded region in Fig.~\ref{fig:fig2}(a)). The fractional frequency noise PSD of the resulting spectrally tailored laser $S_y^\text{S.T.}$ is a combination of the ULE cavity-stabilized laser noise, $S_y^\text{ULE}$, and the Si3 cavity-stabilized comb, $S_y^\text{Si3+comb}$, as follows:
\begin{equation}\label{eqn:stability_transfer}
    S_y^\text{S.T.}=\left|\frac{1}{1+G}\right|^2 S_y^\text{ULE} + \left|\frac{G}{1+G}\right|^2 S_y^\text{Si3+comb},
\end{equation}
where $G$ is the complex transfer function for this phase-locked loop, encompassing the effect of the loop filter and actuator (see Appendix \ref{Append:stability transfer}). Qualitatively, at frequencies much lower than the unity gain frequency of $G$, i.e.~$|G|\gg1$, $S_y^\text{S.T.}$ mostly follows $S_y^\text{Si3+comb}$. On the other hand, at frequencies much higher than the unity gain frequency, i.e.~$|G|\ll1$, $S_y^\text{S.T.}$ will follow $S_y^\text{ULE}$. \textcolor{black}{While the crossover of the frequency noise of ULE cavity-stabilized laser and silicon cavity-stabilized comb occurs at $\sim$\SI{100}{Hz}, we chose the bandwidth of phase locking the ULE cavity stabilized laser to the silicon cavity stabilized comb to be around \SI{500}{Hz}. This is due to a tradeoff between laser noise performance and daily operational robustness of the system. Our ULE cavity is more susceptible to human activity-related perturbations. Thus, a higher lock bandwidth maintains the robustness of the system.}

This spectrally tailored laser features optimized spectral characteristics but its available power is only about \SI{10}{mW}. To increase the available power, a high-power \SI{698}{nm} fiber laser based on sum-frequency generation from \SI{1064}{nm} and \SI{2030}{nm} with a maximum output of \SI{4}{W} (\mbox{PRECILASER} FL-SF-698-4-CW) is phase-locked to the low-power spectrally tailored laser.  This high-power fiber laser has a free-running linewidth of \SI{5.4}{kHz} determined from its measured frequency noise PSD, and thus a \SI{400}{kHz} bandwidth for the phase-locked loop is sufficiently high to tightly transfer the optical phase from the ECDL to the fiber laser. \textcolor{black}{Note that we typically operate this fiber laser at an output power of \SI{1}{W}, limited by the power-handling capacity of our fiber tips. Based on measurements of the in-loop noise of the relevant phase-locked loops, we expect identical noise performance when the laser is operated at the full \SI{4}{W} output power.}

Throughout the whole laser stabilization process, light needs to be delivered to different locations via \textcolor{black}{15- and 20-meter-long} fiber links and a fiber noise cancellation technique is employed to stabilize the path length noise introduced by these fiber links~\cite{Ma1994FNC}. The fiber noise cancellation loops involved here are based on feedback and the sum of their in-loop noise contributions is the ``FNC noise'' in Fig.~\ref{fig:fig2}(b).

The final fractional frequency noise PSD of the high-power spectrally tailored laser delivered to the atoms is presented in the green dashed line in Fig.~\ref{fig:fig2}. It follows the noise performance of the Si3 cavity-stabilized comb within \SI{500}{Hz}. From \SI{500}{Hz} to \SI{10}{kHz}, it gradually follows the noise of the ULE cavity-stabilized laser. At frequencies above $\SI{10}{kHz}$, the FNC noise becomes the dominant noise source. 

\section{\label{sec: on-site frequency noise measurement pulses}Atom-based on-site laser frequency noise measurement}
Despite all the detailed design considerations for laser noise reduction in the previous section, any uncanceled variation of the optical path length delivering light to atoms will add phase noise, reducing the rotation fidelity of the atomic state. Therefore, performing laser frequency noise measurements at the site of the atoms is highly desirable. This can be achieved by utilizing the atomic transition itself as a spectrum analyzer~\cite{bishof2013optical} in which the frequency noise sensitivity is manipulated via a carefully designed pulse sequence~\cite{kolkowitz2016gravitational}, while simultaneously suppressing effects related to laser intensity noise. \textcolor{black}{This follows the general idea of using dynamical decoupling pulses to either eliminate or analyze frequency/phase noise in quantum sensors similar to a lock-in measurement}~\cite{Bollinger2009,Ozeri2011,Oliver2011,Graham2016PRD}. 

For any pulse sequence rotating the final state close to the equator of the Bloch sphere, the single-sided laser frequency noise power spectral density (PSD) $S_{\nu}$ gives rise to a variance of the final excitation probability
\begin{equation}\label{eq:sigma_pe2}
    \sigma^2_{pe} = \pi^2\int^{\infty}_{0} df S_\nu (f)|R(f)|^2.
\end{equation}
Here, $f$ is the Fourier frequency and $R(f)$ is the Fourier transform of the time domain phase sensitivity function $r(t)$ of the final observable~\cite{santarelli1998frequency}: 
\begin{equation}\label{eq:r(t)}
    r(t) = 2\frac{\partial p_e}{\partial \delta\phi}.
\end{equation}
This function measures how much an infinitesimal phase step $\delta \phi$ at time $t$ during the pulse sequence affects the final excitation fraction $p_e$. 

\begin{figure*}[hbtp]
    \centering
    \includegraphics[width=\textwidth]{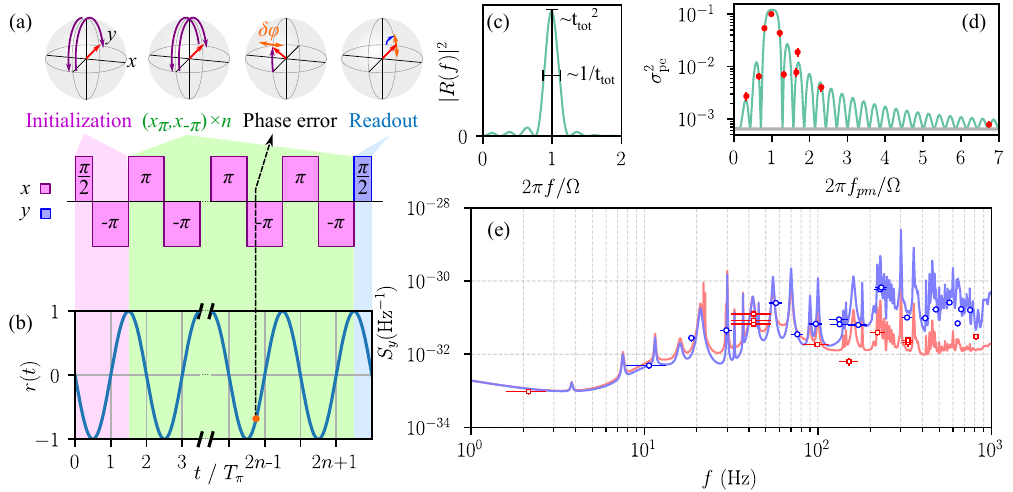}
    \caption{
        \textbf{Phase noise measurement utilizing the atoms.}
        (a) Pulse sequence used to measure the laser phase noise and simultaneously reject pulse area errors. This is achieved by utilizing a maximally phase-sensitive Ramsey sequence which contains continuous back-and-forth rotation of the state vector during the ``dark'' time.
        Atoms prepared in $\ket{e}$ (``north'' pole) are initialized to the equator by pulses around $x$. Then, the atoms are flipped from one side of the equator to the other an integer number $n$ times, canceling pulse area errors by reversing the direction of rotation. A final $\pi/2$ pulse around the $y$-axis maps phase errors $\delta\varphi$ accumulated during the sequence to measurable changes in excitation probability $\delta p_e$.
        (b) The time-domain phase-sensitivity $r(t)$ of the excitation probability is minimal when the state vector points to a pole of the Bloch sphere and maximal at the equator. By reversing the rotation direction exactly on the equator, the oscillation of $r(t)$ at Rabi frequency $\Omega$ is not affected.
        (c) The frequency-domain sensitivity $|R(f)|^2$ (plotted in linear scale here) is peaked around $\Omega/(2\pi)$ and its full width at half maximum is $\sim1/t_\text{tot}$, where $t_\text{tot}$ is the duration of the full pulse sequence. The sensitivity peak amplitude at $\Omega/(2\pi)$ grows with $t_\text{tot}^2$. In order to reject $1/f$ laser noise at low frequencies, a $-\pi$ pulse is inserted in the initialization; this cancels the integral of $r(t)$ and pulls $|R(f)|^2$ to zero at $f=0$.
        (d) Experimental confirmation of the pulse sensitivity function.
        The clock laser is phase-modulated at fixed amplitude $\beta_\text{pm}$ for different modulation frequencies $f_\text{pm}$. Due to the frequency-dependence of the modulation spectrum $S_\nu$, the symmetry of the atomic variance $\sigma_\text{pe}^2$ is reversed with respect to the sensitivity shown in (c). The measured variance values $\sigma_{pe}^2$ (red points\textcolor{black}{, error bars represent the $1\sigma$ error obtained from jackknife resampling}) fit well to the noise expected from the injected incoherent modulation (green line; $\chi^2=$1.540), when adding the unmodulated laser noise (gray line) in quadrature. In this experiment the $-\pi$ pulse in the initialization was omitted.
        (e) Applying this measurement sequence at different Rabi frequencies $\Omega$ allows the extraction of laser phase noise around $\Omega/(2\pi)$. Horizontal bars show the $\pm1/(2t_\text{tot})$ bandwidth of each measurement \textcolor{black}{and vertical bars represent the $1\sigma$ error estimated from the measured variances via jackknife resampling}. By assuming  approximately constant laser noise $S_y$ within this bandwidth around $\Omega/(2\pi)$ and integrating over the sensitivity $|R(f)|^2$, we convert the measured variance to fractional frequency noise. \textcolor{black}{The results under different laser locking conditions are compared against the correspondingly expected laser noise. The blue data points are to be compared with the blue curve, which is identical to the green dashed line in Fig.~\ref{fig:fig2}(b). Red data points are to be compared with the red curve, representing the expected laser noise when a \SI{100}{Hz} bandwidth is applied in the phase lock between the ULE cavity-stabilized laser and the silicon cavity-stabilized comb. For the measurement with \SI{100}{Hz} bandwidth, care was taken to minimize human activity around the ULE cavity for the duration of the experiment.}
        Vertical error bars denote the $1\sigma$ statistical measurement uncertainty.
        }
    \label{fig:fig3}
\end{figure*}

In reality, both laser frequency and intensity noise cause fluctuations of the excitation fraction and this convolution therefore bias the frequency noise measurement. To mitigate this complication, we design a pulse sequence with suppressed sensitivity to intensity noise while maintaining sensitivity to frequency noise (see Fig.~\ref{fig:fig3}(a)). The goal of the pulse sequence is to achieve a maximum sensitivity at a specific Fourier frequency, while allowing this target frequency to be tunable to probe noise over a frequency range of interest. Meanwhile, intensity noise is canceled by the action of neighboring pulses.  

The main temporal structure of the pulse sequence resembles Ramsey spectroscopy: first, the phase sensitivity of the Bloch vector is maximized by rotating it to the equator and mapping the final azimuthal direction to the $z$-axis by adding a rotation around the other transverse axis. In the frame co-rotating with the laser frequency, optical phase noise during the Ramsey ``dark'' time will lead to fluctuations of the azimuthal Bloch vector orientation. These fluctuations may be modeled as a collection of instantaneous phase jumps $\delta\varphi$, which become measurable as $\delta p_e$ after the final readout pulse. To reach the goal of obtaining a singly peaked sensitivity function $|R(f)|^2$, we require $r(t)$ to be oscillating at a constant frequency (see Fig.~\ref{fig:fig3}(b,c)). This would ordinarily correspond to a continuous Rabi oscillation, and thus a smoothly modulated phase sensitivity in time, during the Ramsey ``dark'' time.

To reject variations of Rabi frequency ($\Omega$) due to intensity noise, we modify the constant Rabi drive sequence to one that contains a periodic switching of the rotation direction. This improved sequence prevents an accumulation of pulse area errors over time, making this scheme robust against laser intensity noise. Here, it is important to perform the switch between rotation directions when the state vector reaches the equator of the Bloch sphere, where the phase sensitivity is in one of its extrema. Changing the rotation direction at this point does not change $r(t)$ compared to a continuous Rabi drive without flipping the rotation axis. In the pulse sequence, this is achieved by using a series of repeated $(x_\pi, x_{-\pi})$ rotations, where $x_\pi$ denotes a $\pi$ rotation around the $x$-axis of the Bloch sphere and $x_{-\pi}$ denotes a $-\pi$ rotation around the $x$-axis. Finally, we also insert an additional $x_{-\pi}$ pulse in the initialization section in order to suppress the sensitivity at $f=0$, rejecting detrimental effects from small detunings of the laser frequency from atomic resonance. The sensitivity function $r(t)$ of this full sequence is a windowed sine wave, resulting in a frequency-domain sensitivity $|R(f)|^2$ with a singly peaked maximum at $\Omega/(2\pi)$.

In order to experimentally verify the designed sensitivity function, we inject a controlled amount of laser noise and measure the resulting variance $\sigma_\text{pe}^2$.
The laser frequency is phase modulated at a fixed amplitude $\beta_\text{pm}$ and frequency $f_\text{pm}$. To obtain good signal-to-noise ratio at frequencies far from the sensitivity peak, a sufficiently large value of $\beta_\text{pm}=\SI{10}{\degree}$ is selected. The experimental cycle is chosen to be asynchronous with this modulation and the relative jitter between the phase modulation and the pulse sequence effectively randomizes the phase modulation. The measurement sequence applied here contains $n=3$ blocks of $(x_\pi,x_{-\pi})$ pulses and omits the $x_{-\pi}$ pulse in the initialization.

\textcolor{black}{The contribution to the single-sided laser frequency PSD originating from the coherent modulation tone is}
\begin{equation}\label{eq:PM_PSD}
    S_{\nu}^\text{inj}(f) = \frac{\beta_\text{pm}^2}{2}\delta(f-f_\text{pm})f^2,
\end{equation}
where $\delta(f)$ denotes the Dirac delta function. The tunability of the phase modulation allows probing the atomic sensitivity function without having to change other parameters, e.g.~$\Omega$. Therefore, assuming the intrinsic laser noise $S_\nu^\text{intr}$ is stationary throughout the experiment, the sensitivity function can be characterized without knowing the structure of the laser noise. The laser contribution can be separately quantified from a measurement at a fixed value of $\Omega$ without modulation applied (horizontal gray line in Fig.~\ref{fig:fig3}(d)). \textcolor{black}{These two independent contributions add to the full laser frequency PSD as $S_{\nu}^\text{tot} = S_\nu^\text{intr} + S_\nu^\text{inj}$. We compare the resulting signal} expected from the predicted sensitivity to the experimental results (green line and red points in Fig.~\ref{fig:fig3}(d)). Here, we use Eqs.~(\ref{eq:sigma_pe2}) and (\ref{eq:PM_PSD}) to convert PSD into $\sigma_\text{pe}^2$, and we also take into account the saturation effect in $\sigma_\text{pe}^2$ above $\sim0.03$ due to the curvature of the Bloch sphere (see Appendix \ref{sec:saturation_of_fluctuations}).

After certifying the sensitivity function shown in Figs.~$\ref{fig:fig3}$(b-c) we now apply this tunable atomic spectrum analyzer to probe and verify the laser noise spectrum. For this, we repeatedly apply the pulse sequence shown in Fig.~\ref{fig:fig3}(a) with different Rabi frequencies and numbers $n$ of $(x_\pi,x_{-\pi})$ segments to tune the center frequency and bandwidth of the sensitivity function to probe the laser noise. Here, the center frequency is roughly given by $\Omega/(2\pi)$ and the bandwidth by the inverse of the total pulse duration $t_\text{tot}=2(n+1)T_\pi$, where $T_\pi$ is the $\pi$ pulse duration. The peak of the sensitivity function $|R(f)|^2 $ scales as $\sim t_\text{tot}^2$, which results in a $t_{tot}$ scaling for the variance of the final excited fraction.
By integrating the laser PSD within the sensitivity bandwidth and using it to represent an average value over the probe window, we can move it out of the integral in Eq.~(\ref{eq:sigma_pe2}) and therefore extract a value for $S_\nu(f)$ from the measured variance $\sigma_\text{pe}^2$. \textcolor{black}{We perform this atom-based frequency noise measurement for the experimentally achievable Rabi frequencies to confirm our laser noise model.} \textcolor{black}{Figure~\ref{fig:fig3}(e) compares these values (horizontal bars indicating the $1/t_\text{tot}$ bandwidth) with the laser noise model under different conditions, finding excellent agreement.}

\textcolor{black}{We note another complementary atom-based technique to measure laser frequency noise, demonstrated in Ref.~\cite{Krinner:24}, which involves off-resonantly driving a two-level system and analyzing its response in the incoherent regime. In this approach, the lowest detectable Fourier frequency is limited to approximately half the Rabi frequency, while the highest detectable frequency is, in principle, unbounded—as long as the signal remains above the noise floor. In contrast, our method is limited on the high end by the maximum achievable Rabi frequency but can access much lower Fourier frequencies. Therefore, the two techniques are complementary in terms of the spectral ranges they probe.}

\section{\label{sec: RB}Randomized benchmarking}
\begin{figure}[hbtp!]
    \centering
    \includegraphics{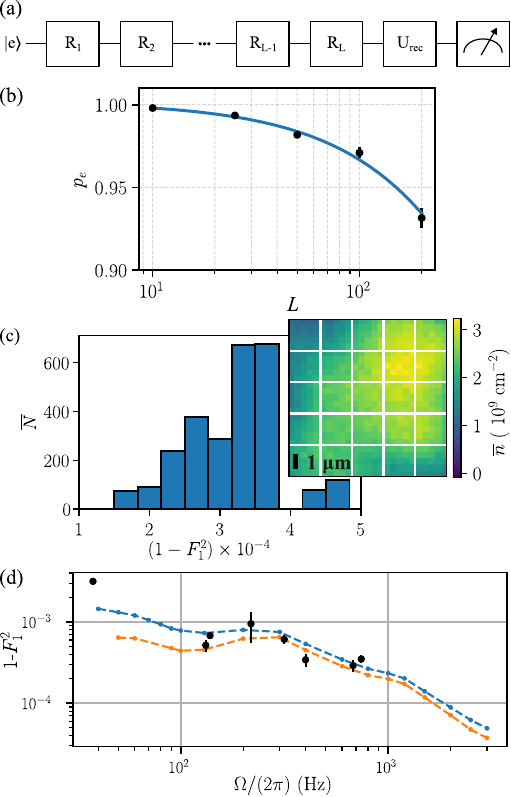}
    \caption{\textbf{Randomized benchmarking performance.}
    (a) Randomized benchmarking is performed by applying a string of $L$ randomly selected single-qubit Clifford gates $R_c$ to atoms initialized to state $\ket{e}$. Finally, a recovery gate $U_\text{rec}$ rotates the state vector back to its starting position on the Bloch sphere.
    (b) Decay of $p_e$ versus $L$ with pulse Rabi frequency $\Omega=2\pi\times\SI{741.9(5)}{Hz}$ for $\sim3000$ atoms within a \SI{10.3}{\micro \meter} $\times$ \SI{10.3}{\micro\meter} (25$\,$pixels $\times$ 25$\,$pixels) region of interest. For each $L$ the measurement of $p_e$ is repeated 50 times (average shown as black points\textcolor{black}{, and vertical error bars represent the standard error of the average excitation fraction}). The blue line shows a fit of Eq.~(\ref{eq:fidelity_fit_decay}) for all atoms in the region of interest and a single Clifford gate fidelity of $F_1^2 = 0.99964(3)$ is derived from the fit result.
    (c) By sub-dividing the region of interest into superpixels of the same size, we evaluate the inhomogeneity of the gate fidelity. The histogram shows the distribution of the measured fidelities with respect to the average atom number $\overline{N}$ in the corresponding sets of superpixels. The inset shows the averaged absorption image of the atomic column density distribution $\overline{n}$ along with the sub-division of the region of interest.
    (d) Averaged single-qubit infidelities $1-F_1^2$ extracted from fits \textcolor{black}{(black points, error bars represent the $1\sigma$ statistical error)} analogous to \textcolor{black}{(b)} are compared to a simulation of the randomized benchmarking for different Rabi frequencies. The simulation accounts for the frequency noise spectrum shown in Fig.~\ref{fig:fig3}(e). The blue dashed line includes further a Raman-scattering-limited excited state decay rate of \SI{0.077(5)}{s^{-1}} and a measured decoherence rate of \SI{0.143(4)}{s^{-1}}; and the orange dashed line does not.
    }
    \label{fig:fig4}
\end{figure}

With this spectrally tailored laser, how well can we control the optical qubit? To quantify the fidelity of quantum state engineering, we perform randomized benchmarking based on single-qubit Clifford gates to measure the average gate fidelity~\cite{knill2008randomized,Easwar2011Benchmarking, Xia2015PRL}. Fig.~\ref{fig:fig4}(a) shows a schematic of the pulse sequence for the randomized benchmarking. We initialize atoms in the excited state $\ket{e}$ and apply a sequence of $L$ randomly selected elements $R_i$ of the single-qubit Clifford group $C_1$ with uniform probability. 
These are realized by a combination of $\pi$ and $\pi/2$ rotations around the $\pm x$ and $\pm y$ axes as tabulated in Table \ref{tab:Clifford group}, with an average pulse area of $9/8\,\pi$ per Clifford gate.
A recovery pulse $U_{rec}\in C_1$ is applied after the random Clifford gate sequence. For a given Clifford gate sequence, $U_{rec}$ is randomly chosen from a set of four Clifford gates that will restore the Bloch vector to $\ket{e}$ under ideal operation. By applying the clock laser to all atoms simultaneously and measuring the populations in $\ket{g}$ and $\ket{e}$, we obtain a value for $p_e$ in each realization of the experiment.

The gate fidelity averaged over all single-qubit Clifford gates can be extracted by measuring how the probability of reaching the correct output state, i.e.~$p_e$, decays when $L$ increases.
This probability is captured by the fidelity $F^2(\rho_0,\rho)$ between the pure initial state specified by the density matrix $\rho_0=\ket{e}\bra{e}$ and the final state $\rho$,
\begin{equation}
    F^2(\rho, \rho_0) = \left(\text{Tr}\left[\sqrt{\sqrt{\rho}\rho_0\sqrt{\rho}}\right]\right)^2\,.
\end{equation}
Applying a purely depolarizing model that assumes the only effect of gate errors to be an increase of the mixing of $\rho$, the fidelity of the full benchmarking sequence is given by~\cite{Xia2015PRL}
\begin{equation}\label{eq:fidelity_fit_decay}
    F_L^2 = p_e = \frac{1}{2}+\frac{1}{2}(1-d_\text{SPAM}
)(1-d)^L,
\end{equation}
where $d$ is the depolarization probability for a single pulse and $d_\text{SPAM}$ is the state preparation and measurement error. The average fidelity of a single Clifford gate is $F_1^2 = 1 -d/2$.

Figure~\ref{fig:fig4}(b) shows randomized benchmarking results for gates at Rabi frequency $\Omega=2\pi \times \SI{741.9(5)}{Hz}$ for 3000 atoms. We measure a single-qubit gate fidelity $F_1^2 = 0.99964(3)$ averaged over the entire cloud and all 24 single-qubit Clifford gates. The distribution of fidelities across the atomic cloud is narrowly peaked around this value, as shown in the histogram in Fig.~\ref{fig:fig4}(c), which is obtained by sub-dividing the images into a grid (cf.~inset). This indicates an excellent uniformity of high-fidelity state manipulation across the entire sample of lattice-confined atoms. 

To study how laser frequency noise impacts the fidelity of pulses, randomized benchmarking measurements at different Rabi frequencies are performed. The experiment results are plotted with black dots in Fig.~\ref{fig:fig4}(d) and compared to a numerical simulation (blue dashed line) based primarily on the laser frequency noise PSD, showing good agreement between the two. Contributions from atomic decay and decoherence are small and only relevant at time scales beyond \SI{1}{s}~\cite{Hutson2019PRLshallow}. Small deviations in certain Rabi frequency ranges arise from peaks in the PSD from the vibration of the cavity that may drift over time within limited ranges. 
Towards higher Rabi frequencies we observe an increasing fidelity, consistent with the understanding that shorter pulses should lead to less sensitivity to fluctuations.

\section{Outlook}
Development of highly stable lasers provides critically important capabilities for the continued advance of quantum science and technology. The realization of a spectrally tailored high-power laser for driving optical transitions highlights an example where systematic considerations of laser spectral noise are integrated closely with an atom-based quantum system for performance characterization and verification. We hope this case study serves a useful technical purpose for other researchers interested in developing highly stabilized lasers for their specific applications. Particularly, the reported system has already demonstrated a highly desired property of achieving a record fidelity for manipulating a large and scalable number of optical qubits.  

To further improve the fidelity and robustness of optical manipulations based on a sequence of laser pulses, we can consider two approaches. The first is to engineer pulses that are less sensitive to uncontrolled experiment parameters, e.g.~employment of composite pulses~\cite{PRA2021composite_gates}, pulse shaping in amplitude and phase~\cite{PulseShaping2007PRA,PRA2008PulseShapingYbIon} or utilization of optimal control algorithm to design pulses~\cite{KHANEJA2005GRAPE}. The other is to further reduce the relevant high-frequency noise by developing even faster frequency transducers to advance servo bandwidth to increasingly higher values. Feedforward techniques~\cite{li2022activefeedforward,chao2024robustfeedforward} can also be employed to complement servo loops to mitigate frequency noise beyond the bandwidth of available feedback techniques. \textcolor{black}{Loading more atoms with low entropy, together with improved beam shaping, such as using a top-hat beam for the clock laser, can improve the number of qubits that can be homogeneously addressed simultaneously.}

%\textcolor{blue}{The development of such laser systems also supports broader goals beyond quantum information. A key advantage of atomic qubits lies in their strong connection to fundamental physics. The long coherence times and spectral resolution enabled by this platform allow exploration of phenomena ranging from collective light-matter interactions (e.g., superradiance and collective Lamb shifts) to relativistic effects such as gravitational redshifts on microscropic scales. As quantum simulation efforts progress, the ability to precisely engineer Floquet Hamiltonians and probe complex many-body spin models (e.g., the XYZ model) will benefit significantly from high-fidelity optical control. Our system thus not only advances the technical frontier for quantum computing but also creates new opportunities for foundational studies in quantum metrology and beyond.}

The vastly improved fidelity for driving optical transitions in scalable and highly coherent atomic/molecular systems will improve atom-based quantum computing. Reduction of quantum state preparation error is also in strong demand for realizing metrologically useful spin squeezing. With the capability of applying hundreds of pulses to drive an optical transition without significantly losing contrast,  we can Floquet engineer an effective Hamiltonian of a many-body system to decouple unwanted interactions or to study novel emergent phenomena using specially designed pulses~\cite{choi2020robust}. This method has already found a widespread use in the microwave/RF domain due to the relative ease for high-fidelity RF pulses~\cite{Souza2011,Zhou2023,miller2024KRbFluoquet}. However, because of the limited laser stability, this approach has not achieved similar popularity in the optical domain. We hope to change this landscape with the development reported here. Particularly for our 3D optical lattice clock platform, existing interactions such as collective dipolar couplings~\cite{hutson2024observation} and superexchange interactions~\cite{milner2024coherent} can be engineered to realize longer atomic coherence and metrologically useful entangled states~\cite{Mamaev2024superexchange}.

\textbf{Acknowledgment} We are grateful to A.~Aeppli, Z.~Hu, J.~Hur, D.~Kedar, Y.~Lee, M.~Miklos, J.~M.~Robinson, Y.~M.~Tso, W.~Warfield, Z.~Yao, and J.~L.~Hall for technical discussions and assistance. We thank N.~D.~Oppong, E.~Bohr, J.~P.~Covey, B.~Bloom, and J. Home for careful reading of the manuscript and for providing insightful comments. Funding for this work is provided primarily by DOE Center of Quantum System Accelerator, and also by V. Bush Fellowship, NSF QLCI OMA-2016244, NSF JILA-PFC PHY-2317149, and NIST. S.~L.~acknowledges funding from the Alexander von Humboldt Foundation and B.~L.~from the Lindemann Trust.

\textbf{Author contributions}
All authors contributed to the stable laser development. The atom-based measurement was carried out by L.Y., S.L., W.R.M., M.N.F., and J.Y. on the JILA Sr2 quantum gas-3D lattice clock platform. All authors contributed to data analysis and writing of the paper. 

\textbf{Data availability}

All data is available from the corresponding
authors on reasonable request.

\textbf{Competing interests}

The authors declare no competing interests.

\appendix

\section{\label{Append_A: laser nosie details}Details of laser noise modeling}
In this section we provide further technical details about the characterization and modeling of laser noise components when designing the spectrally tailored high-power laser described in Sec.~\ref{sec:laser frequency noise reduction}. 

\subsection{\label{Append_A_1:Si}Si3 cavity noise characterization}

\begin{figure}
    \centering
    \includegraphics[width=\columnwidth]{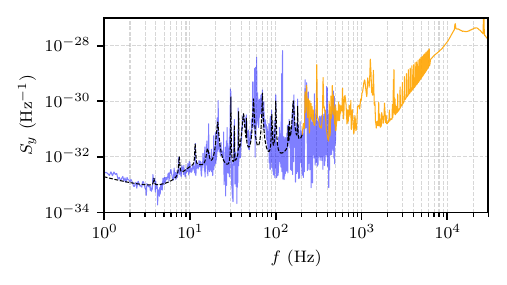}
    \caption{\textcolor{black}{\textbf{Evaluation of Si3 cavity-stabilized laser noise.} The blue and orange traces show the cross-spectral measurements using a counter (with a Nyquist frequency of \SI{500}{Hz}) and a tracking filter (with a \SI{10}{kHz} bandwidth; above this frequency, the measured noise represents only a lower bound). The black dashed line represents the fractional frequency noise power spectral density model of the Si3 cavity below \SI{200}{Hz}}.}
    \label{fig:A1:Si_model}
\end{figure}

The noise of the \SI{1542}{nm} laser frequency stabilized to the Si3 cavity is evaluated with a cross-correlation measurement~\cite{rubiola2000correlation} against two other independent frequency-stabilized lasers: a cryogenic \SI{6}{cm} silicon cavity operating at \SI{17}{K} and a room temperature \SI{40}{cm} ULE cavity. The cross-correlation measurement is based on the recorded heterodyne beat frequencies among these three cavities, and a frequency comb is needed to bridge the spectral gap between \SI{698}{nm} of ULE and \SI{1542}{nm} of silicon. Specifically, the cross-correlation of the optical beats between Si3 and the other two cavities provides the frequency noise power spectral density of the Si3 cavity-stabilized laser. We perform the cross-correlation measurement using two different approaches. First, we use a frequency counter to simultaneously record the frequency of the two optical beats. The counter has a minimum gate time of \SI{1}{ms}, which limits the sampling frequency to about \SI{500}{Hz}. Therefore, we discard information above \SI{500}{Hz} from the counter-based measurement. This counter-based cross-spectral density evaluation is represented by the \textcolor{black}{blue} trace in Fig.~\ref{fig:A1:Si_model}. In order to retrieve high-frequency noise of the Si3 cavity-stabilized laser, we use two phase-locked tracking filters to record the phase information of the two relevant optical beats. 
The tracking filters have \SI{10}{kHz} bandwidth and they are limited by their intrinsic noise below \SI{200}{Hz}. Therefore, the tracking filter-based cross-correlation measurement provides information of the Si3 cavity-stabilized laser from \SI{200}{Hz} to \SI{10}{kHz} (shown as the \textcolor{black}{orange} trace in Fig.~\ref{fig:A1:Si_model}). Measurements based on these two different methods are consistent with each other in their frequency overlapped region. 
The additive noise of the frequency comb is inferred from our previous work~\cite{oelker2019demonstration} and grants a negligible contribution relative to the Si3 noise.

\textcolor{black}{We proceed by constructing a model of the Si3 cavity-stabilized comb from these two separate noise measurements. At Fourier frequencies above \SI{200}{Hz}, we use a tracking filter-based cross spectral density measurement. Below \SI{200}{Hz} we use the counter-based cross-spectral density measurement and fit the measured noise spectrum with the following analytical form:}
%Based on the counter-based result of the cross-spectral density measurement, we use the following function to model the noise of Si3 cavity-stabilized comb for noise below \SI{200}{Hz}:
\textcolor{black}{\begin{equation}
    S^{\text{Si3+comb}}_{y}(f) = \frac{h_{-1}}{f} + h_0 + h_2 f^2 + \sum_{i=1}^{19}\frac{a_i^\text{Si3+comb}f^2}{1+\left(\frac{f-f_i^\text{Si3+comb}}{\Gamma_i^\text{Si3+comb}/2}\right)^2}.
\end{equation}}
Here we include a thermal frequency flicker noise term ($h_{-1}$ = 1.5 $\times$ 10$^{-33}$), a white frequency noise term ($h_0$ = 4.0 $\times$ 10$^{-34}$ Hz$^{-1}$), a white phase noise term ($h_2$ = 4.3 $\times$ 10$^{-37}$ Hz$^{-3}$), and a series of \textcolor{black}{mechanical }resonant features. Ref.~\cite{oelker2019demonstration} provides similar information for $f <$ \SI{100}{Hz}.
\textcolor{black}{The \SI{60}{Hz} and \SI{120}{Hz} peaks visible in the counter measurement arise from electronic pickup, as confirmed by the fact that they do not appear in the direct atomic spectral analysis, and are therefore not represented in the noise model.}

As for the noise at $f >$ \SI{200}{Hz}, we directly interpolate the tracking filter-based cross spectral density measurement result to describe the noise of Si3 cavity-stabilized comb.
The model is plotted as the black dashed line in Fig.~\ref{fig:A1:Si_model} to compare against the cross-correlation measurement results.

\subsection{\label{Append:A2:ULE caivty}High-bandwidth PDH lock to a ULE cavity}
\begin{figure}
    \centering
    \includegraphics[width=\columnwidth]{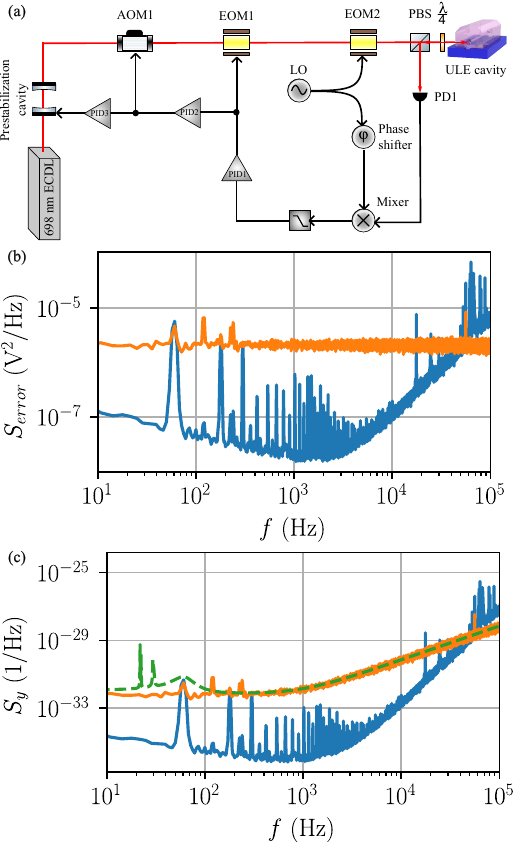}
    \caption{\textbf{Details of the ULE cavity-referenced laser.} (a) Schematic of the high-bandwidth PDH lock to the ULE cavity. (b) PDH lock in-loop error signal PSD (blue) and detector noise floor PSD (orange). (c) Fractional frequency noise level corresponding to the data shown in (b). Additionally, the ULE cavity noise model is shown as green dashed line.}
    \label{fig:A2_ULE_noise_detial}
\end{figure}

Instead of directly locking the \SI{698}{nm} external cavity diode laser (ECDL) to the ULE cavity, our setup features an additional prestabilization stage with an electro-optic modulator (EOM) as a fast actuator (see Fig.~\ref{fig:A2_ULE_noise_detial}(a) for a schematic). The ECDL is prestabilized to a medium finesse cavity (finesse of 50,000, linewidth of \SI{100}{kHz}) with an auxiliary PDH loop (not shown in the schematic). The transmission light of this prestabilization cavity is the starting point for the rest of the high-bandwidth PDH lock to the narrow linewidth (\SI{2}{kHz}) ULE cavity. A resonant EOM (EOM2) is driven with a \SI{9.2}{MHz} RF signal to generate sidebands on the incident light to the ULE cavity for the PDH lock. A photodiode (PD1) collects the reflected light from the cavity, and its photocurrent is demodulated to generate the PDH error signal. This error signal is passed to a series of proportional-integral-derivative (PID) loop filters and actuators, whose design is inspired by Ref.~\cite{hall1984external,Zhu:93}. PID1 directly takes the PDH error signal as its input, and a low delay time amplification stage drives a broadband EOM (EOM1), which can realize $\approx \SI{1}{MHz}$ bandwidth but with a limited dynamic range for correcting phase errors. To increase the servo range, the output of PID1 serves as the input to PID2, which controls the RF drive frequency to an acousto-optic modulator (AOM1) with a bandwidth of $\approx\SI{100}{kHz}$. Finally, the output of PID2 serves as the input error signal for PID3, which controls the length of the prestabilization cavity with a piezoelectric actuator. This final loop has a few kHz bandwidth but with the largest servo correction dynamics range. The three PID loops are connected in series to prevent partial cancellation of their outputs and still reach an overall bandwidth of \SI{1}{MHz}. \textcolor{black}{Choosing this bandwidth ensures laser noise suppression to the detection noise floor up to Fourier frequencies well above the range of Rabi frequency relevant in our system.}

The frequency noise of a cavity-stabilized laser is determined by the noise of the cavity itself, the in-loop error of the servo loop, and the residual photodetection noise. For our system, the in-loop error is below the detection noise floor up to $\textcolor{black}{\SI{50}{kHz}}$. This means that the actual in-loop noise within $\textcolor{black}{\SI{50}{kHz}}$ is dominated by the white shot noise of PDH photo detection. A cavity serves as a frequency noise discriminator for laser frequency/phase \textcolor{black}{fluctuations} at time scales slower than the cavity lifetime and it turns into a phase noise discriminator for fluctuation faster than the cavity lifetime~\cite{drever1983laser}. Thus, the slope of the PDH frequency discriminator $k(f)$ has a following frequency-dependent form~\cite{simplephasemeasurement2019, nagourney2014quantum}:
\begin{equation}\label{eqn:pdh_transfer}
    k(f) = \frac{k_0}{\sqrt{1+(f/\Delta f_\text{HWHM})^2}},
\end{equation}
where $k_0$ is the slope of the DC PDH error signal, $f$ is the Fourier frequency of laser noise, and $\Delta f_\text{HWHM}$ is the HWHM of the cavity. For the ULE cavity utilized in our setup, $\Delta f_\text{HWHM}$ = 1 kHz~\cite{mattS2012highPrecisionClock}.  In the frequency domain where the in-loop error signal is below the detection noise floor, the photon shot noise of the PDH detector is turned to white frequency noise for $f < \Delta f_\text{HWHM}$ and then white phase noise for $f > \Delta f_\text{HWHM}$. 

The frequency noise of the ULE cavity up to $f$ of \SI{100}{Hz} was previously characterized using atoms, revealing its thermal noise floor and vibration-induced resonance peaks~\cite{bishof2013optical}. With the additional characterization between \SI{100}{Hz} and \SI{100}{kHz},  we can sum all the contributions and reach the following model for the frequency noise of the ULE cavity-stabilized laser:
\textcolor{black}{\begin{equation}
    S_y^{\text{ULE}}(f) = \frac{h_\text{thermal}}{f} + h_\text{white} + h_2^\text{ULE}f^2 + \sum_{i=1}^{3} \frac{a_i^\text{ULE}}{1+\left(\frac{f-f_i^\text{ULE}}{\Gamma_i^\text{ULE}/2}\right)^2}.
\end{equation}}
\textcolor{black}{Here, a thermal flicker noise term ($h_\text{thermal}=8.2\times10^{-33}$), a white frequency noise term ($h_\text{white}=7.1 \times 10^{-33}\,$Hz$^{-1}$), a white phase noise term ($h_2^\text{ULE} = h_\text{white}/(\Delta f_\text{HWHM})^2$) and a series of 3 resonant features are included.} The thermal flicker noise term and resonant features are adopted from Ref.~\cite{bishof2013optical} while the white frequency noise term and the white phase noise term are updated.

Figure~\ref{fig:A2_ULE_noise_detial}(b) displays the PDH in-loop error signal when the laser is locked to the ULE cavity (blue) versus the PDH photo detection noise floor (orange). The photon shot noise is measured when the laser frequency is tuned far away from the cavity resonance and the same amount of optical power is incident on the PDH photodetector as compared to when the laser is locked to the cavity. The in-loop servo error noise is below the photo-detection noise floor for $f <$ \textcolor{black}{\SI{50}{kHz}}. The corresponding laser frequency noise contributions are plotted in Fig.~\ref{fig:A2_ULE_noise_detial}(c) against the model of the ULE cavity-stabilized laser.

\subsection{\label{Append:stability transfer}Si cavity to ULE cavity stability transfer loop}
\begin{figure}
    \centering
    \includegraphics[width=\columnwidth]{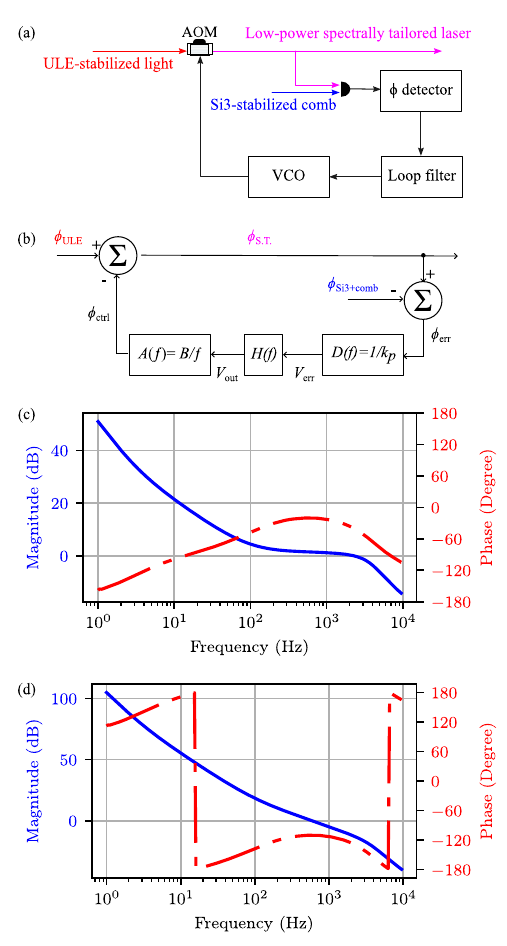}
    \caption{\textbf{Si cavity to ULE cavity stability transfer loop.} (a) Simplified schematic of the stability transfer loop. (b) Block diagram of the stability transfer loop. (c) Bode plot of the loop filter transfer function $H(f)$. (d) Bode plot of the open-loop gain $G(f)$ of the full stability transfer loop.}
    \label{fig:A3_stability_transfer}
\end{figure}

The schematic of the stability transfer loop, which was initially set up for the Sr clock operation, can be found in Ref.~\cite{Lindsay2021thesis}. For this work we use a simplified version shown in Fig.~\ref{fig:A3_stability_transfer}(a) to illustrate the concepts. The ULE cavity-stabilized laser passes through an AOM and the first-order diffracted light is used to beat against the Si3 cavity-stabilized comb light. The phase noise of the beat signal is detected against a stable RF reference via a phase detector, and the derived error signal goes through a loop filter to drive a voltage-controlled oscillator (VCO), which provides the RF drive of the AOM. Figure~\ref{fig:A3_stability_transfer}(b) illustrates the basic model of this phase-locked loop in frequency domain. Here, $\phi_\text{ULE}$ and $\phi_\text{Si3+comb}$ represent the optical phase of the ULE and Si3 cavity-stabilized lasers, respectively.
These are used to derive the spectrally tailored laser phase as
\begin{equation}\label{eq:phi_synth}
    \phi_\text{S.T.}=\phi_\text{ULE}-\phi_\text{ctrl},
\end{equation}
where $\phi_\text{ctrl}$ is the phase correction signal of the AOM derived by the loop. The phase detector converts the phase error $\phi_\text{err}$ to a voltage $V_\text{err}$, which is the input signal of the feedback loop.
For the frequency range relevant to our system the transfer function
\begin{equation}\label{eq:D(f)}
    D(f) = 1/k_p
\end{equation}
of the phase detector is essentially a constant, where $k_p = \SI{11.63}{Rad/V}$ is the conversion factor.
Fig.~\ref{fig:A3_stability_transfer}(c) shows a measured Bode plot of the loop filter transfer function $H(f)$. Its output serves as the control signal for the VCO driving the AOM, which has a frequency-tuning sensitivity to the control voltage of $k_v = 2\pi \times \SI{5.84}{kHz/V}$. The transfer function of this frequency actuator contains an intrinsic 1/$f$ for phase actuation,
\begin{equation}\label{eq:Actuator_transfer_function}
    A(f) = B / f,
\end{equation}
where $B = k_v / (i 2\pi)$ and $i$ is the imaginary unit.
Thus,
\begin{equation}\label{eq:Vout_1}
    V_\text{out} = HV_\text{err}=\frac{H}{k_p}\phi_\text{err}=\frac{H}{k_p}(\phi_\text{S.T.}-\phi_\text{Si3+comb})
\end{equation}
and
\begin{equation}\label{eq:phi_ctrl}
    \phi_\text{ctrl} = AV_\text{out}.
\end{equation}
By combining Eqs.~(\ref{eq:phi_synth}), (\ref{eq:Vout_1}) and (\ref{eq:phi_ctrl}) we obtain
\begin{equation}\label{eq:Vout_2}
    V_\text{out}= \frac{H}{k_p+BH/f}\,(\phi_\text{ULE}-\phi_\text{Si3+comb}).
\end{equation}
Substituting Eqs.~(\ref{eq:phi_ctrl}) and (\ref{eq:Vout_2}) into Eq.~(\ref{eq:phi_synth}), we arrive at
\begin{equation}\label{eq:stability_phi_synth}
    \phi_\text{S.T.} = \frac{1}{1+G}\,\phi_\text{ULE}+\frac{G}{1+G}\,\phi_\text{Si3+comb}
\end{equation}
with the open-loop gain of the stability transfer loop
\begin{equation}\label{eq:G(f)}
    G(f)=\frac{B}{k_p f}H(f).
\end{equation}
The Bode plot of $G(f)$ is shown in Fig.~\ref{fig:A3_stability_transfer}(d). Because the ULE and Si3 noise sources are independent, one obtains the noise power of Eq.~(\ref{eqn:stability_transfer}) by simply squaring the summands in Eq.~(\ref{eq:stability_phi_synth}).

\textcolor{black}{\subsection{Fiber noise cancellation loops}\label{sub_FNC}}

\begin{figure}
    \centering
    \includegraphics[width=\columnwidth]{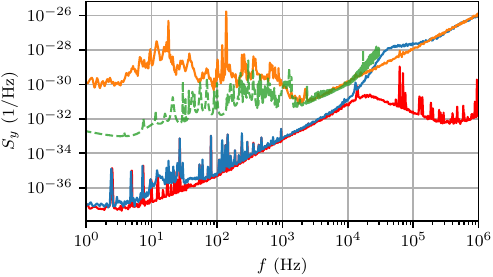}
    \textcolor{black}{\caption{\textbf{FNC loop characterization.} Fractional frequency noise power spectral density $S_y$ of the FNC beat. The orange trace shows the FNC beat when feedback is off, corresponding to twice the fiber-induced path length noise. The blue trace shows $S_y$ of the FNC beat when the FNC loop is active. For reference, we also show the $S_y$ of the FNC loop’s local oscillator (red trace) and the spectrally tailored laser (green trace).}\label{fig:A4:FNC}}   
\end{figure}

\textcolor{black}{
The stabilized laser is distributed to multiple experimental locations via fiber links, with fiber noise cancellation (FNC) loops applied to suppress path length noise introduced by the fibers~\cite{Ma1994FNC}. In Fig.~\ref{fig:A4:FNC}, we present a measurement of the fractional frequency noise power spectral density $S_y$ of the FNC loop for one of the \SI{20}{m} long fibers in our system, compared to that of the spectrally tailored laser (green trace). The orange trace shows the $S_y$ of the FNC beat when the FNC feedback is disabled, representing twice the amount of path length noise introduced by the fiber. Without active cancellation, this fiber-induced noise would severely degrade the performance of the spectrally tailored laser below \SI{1}{kHz}. \textcolor{black}{At high frequencies, the uncancelled fiber noise grows as  $f^2$, consistent with white phase noise. This white phase noise floor within the measurement bandwidth arises from photon shot noise in the optical heterodyne phase detection process.} When the FNC loop is activated, the FNC beat (blue trace) is stabilized to the local oscillator of the FNC loop (red trace), and a servo bump appears around \SI{30}{kHz}.
    }

\textcolor{black}{
We note that the overlap between the orange and green traces in the \SI{2}{kHz} to \SI{10}{kHz} range is coincidental; in this frequency region, the laser noise is dominated by white phase noise from the ULE cavity PDH lock.
}
\vspace{6ex}
\section{Accounting for saturation effects in the fluctuation of the excitation probability}\label{sec:saturation_of_fluctuations}
Eq.~(\ref{eq:sigma_pe2}) is derived in the limit that the fluctuations of the final excitation probability due to laser noise are small. This is the limit where the curvature of the Bloch sphere is negligible. In case of large excursions of the excitation probability away from the equator, $\sigma_{pe}^2$ will not grow linearly with the laser noise, but instead become saturated. The saturation arises from the fact that the maximum deviation of $p_e$ can only be $1/2$, i.e.~when reaching either pole of the Bloch sphere. Thus, to deal with the saturation effect, instead of analyzing $p_e$, we should use the polar angle $\theta$ of the final state on the Bloch sphere as a quantity in the laser noise model.  

For the treatment of laser noise, we have initially performed a linearization of the laser phase fluctuations (cf.~Eq.~(\ref{eq:r(t)})), which also constrains all derived quantities to the linear limit. Eq.~(\ref{eq:sigma_pe2}) is hence applicable to the variance $\sigma_\theta^2$ of $\theta$ and this equivalence holds even beyond the point where $\sigma_{pe}^2$ starts showing saturation effects. Therefore, instead of $\sigma_{pe}^2$, we now utilize $\sigma_\theta^2$ to derive a prediction of the variance of the excitation probability $p_e$, which is valid even when $\sigma_{pe}^2$ saturates.

For this, we introduce a parametrization of the excitation probability $p_e$ in terms of the polar deviation $\tilde{\theta}=\theta-\pi/2$ from the equator,
\begin{equation}\label{eq:pe}
    p_e=\frac{1}{2}-\frac{\sin\tilde{\theta}}{2}.
\end{equation}
In line with our noise models, we assume $\tilde{\theta}$ to be a Gaussian random variable, i.e.~its probability distribution is given by
\begin{equation}\label{eq:p(theta)}
    p(\tilde{\theta})=\frac{1}{\sqrt{2\pi}\sigma_\theta}\exp\left(-\frac{\tilde{\theta}}{2\sigma_\theta^2}\right)
\end{equation}
with variance $\sigma_{\tilde{\theta}}^2=\sigma_\theta^2$ given by Eq.~(\ref{eq:sigma_pe2}). By applying Eqs.~(\ref{eq:pe}) and (\ref{eq:p(theta)}) we obtain expressions for the first and second moment of $p_e$ in dependence of $\tilde{\theta}$ or rather $\sigma_\theta^2$:
\begin{align}
    \begin{split}
        \langle p_e\rangle&=\int_{-\infty}^\infty d\tilde{\theta}\left(\frac{1}{2}-\frac{\sin\tilde{\theta}}{2}\right)p(\tilde{\theta})=\frac{1}{2}, \\
        \langle p_e^2\rangle&=\int_{-\infty}^\infty d\tilde{\theta}\left(\frac{1}{2}-\frac{\sin\tilde{\theta}}{2}\right)^2p(\tilde{\theta}) \\
        &=\frac{1}{4}+\frac{\exp(-\sigma_\theta^2)}{4}\sinh(\sigma_\theta^2).
    \end{split}
\end{align}
This allows expressing the variance $\sigma_\text{pe}^2$ of the excitation probability $p_e$ as
\begin{equation}
    \sigma_\text{pe}^2=\frac{1}{4}\exp(-\sigma_\theta^2)\sinh(\sigma_\theta^2)\,,
\end{equation}
where $\sigma_\theta^2$ is given by Eq.~(\ref{eq:sigma_pe2}). Note that this expression is valid so long as the error $\delta p_e$ of the final excitation probability stays linear with the phase error $\delta\phi$, which remains true for \textcolor{black}{fluctuations} much larger than those leading to a saturation of $\sigma_\text{pe}^2$ due to the curvature of the Bloch sphere.

\section{Numerical simulation of gate fidelity}\label{append:Numerical simulaiton of gate fidelity}
Our experimental realization of each element of single-qubit Clifford group is tabulated in Table~\ref{tab:Clifford group}. To numerically study the averaged single-qubit Clifford gate fidelity, we use the following master equation to simulate the randomized benchmarking experiment:
\begin{equation}\label{eq:master_eq}
    \frac{d\rho}{dt}=-\frac{i}{\hbar}[H,\rho]+\mathcal{L}(\rho).
\end{equation}
Here, $\rho$ is the atomic density matrix, 
\begin{equation}\label{eq:Lindblad}
\begin{split}
\mathcal{L}(\rho)= & \Gamma(\sigma_-\rho\sigma_+ - \frac{1}{2}{\{ \sigma_+ \sigma_- ,\rho\} }) \\
&+ \frac{\gamma_c}{2}(\sigma_z\rho\sigma_z^\dagger - \frac{1}{2}{\{ \sigma_z^\dagger \sigma_z ,\rho\} })
\end{split}
\end{equation}
is the Lindblad dissipator that captures the atom-atom decoherence effect, and 
\begin{equation}\label{eqn:2_level_Hamiltonian}
    H = \frac{\hbar\delta}{2}\sigma_z + \frac{\hbar\Omega}{2}[\cos{(\phi_0+\delta\phi)}\sigma_x+ \sin{(\phi_0 + \delta\phi)}\sigma_y] 
\end{equation}
is the Hamiltonian (with rotating-wave approximation applied) that captures the dynamics of the driven two-level system in the presence of laser noise. Here, $\Gamma$ is the excited state decay rate, $\gamma_c$ the additional decoherence rate, $\delta$ the laser detuning, $\Omega$ the Rabi frequency, $\phi_0$ the laser phase defining the rotational axis and $\delta\phi$ captures the laser noise.

In the limit of $\Omega=0$ and $\delta=0$, Eq.~(\ref{eq:master_eq}) will be reduced to the following form:
\begin{equation}
\begin{aligned}
    \partial_t \rho_{ee} &= -\Gamma \rho_{ee}, \\
    \partial_t \rho_{gg} &= +\Gamma \rho_{ee}, \\
    \partial_t \rho_{ge} &= -\gamma_\perp \rho_{ge}, \\
    \partial_t \rho_{eg} &= -\gamma_\perp \rho_{eg},
\end{aligned}
\label{eq:density_matrix}
\end{equation}
where one can easily find that $\Gamma$ is the excited state population decay rate and $\gamma_\perp=\frac{\Gamma}{2}+\gamma_c$ is the total decoherence rate.
For this work, the operational lattice depth is $(V_x, V_y, V_z) = (45, 44, 41)E_r$. This is a regime where $\Gamma$ is limited by Raman scattering of \SI{813}{nm} lattice photons~\cite{Hutson2019PRLshallow}. Based on the operational lattice depth, we use $\Gamma=\SI{0.077(5)}{\per\second}$ for the numerical modeling. 
We experimentally measure the atomic coherence time of \SI{7.0(2)}{s} via extracting the decay time constant of Ramsey contrast using imaging spectroscopy~\cite{marti2018imaging}, i.e. $\gamma_\perp = \SI{0.143(4)}{\per\second}$. Therefore, the parameter $\gamma_c$ in Eq.~(\ref{eq:Lindblad}) is $\gamma_c = \gamma_\perp -\Gamma/2= \SI{0.105(5)}{\per\second}$.

$\phi_0$ is a deterministic phase that defines the rotation axis. $\phi_0 = -\frac{\pi}{2}, 0, \frac{\pi}{2}, \pi$ corresponds to the rotation along $-y$, $x$, $y$, $-x$ axis. And $\delta\phi$ in Eq.(\ref{eqn:2_level_Hamiltonian}) is the stochastic phase fluctuation due to laser phase noise. For each run of the simulation, a time trace of $\delta\phi$ is generated based on the single-sided phase noise power spectral density $S_{\phi}$ of the laser~\cite{li2022activefeedforward, jiang2023sensitivity}:
\begin{equation}\label{eq:phi(t)}
    \delta\phi(t) =\sum_f\sqrt{2S_\phi(f)\Delta f}\cos{(2\pi ft + \phi_f)},
\end{equation}
where $\phi_f$ is the random phase offset of Fourier frequency $f$ and $\Delta f$ is the frequency resolution of $S_{\phi}(f)$. $S_{\phi}(f)$ is related to the fractional frequency noise of laser $S_y(f)$ discussed in Sec.~\ref{sec:laser frequency noise reduction} by $S_{\phi}(f)=S_y(f)\nu_0^2/f^2$, where $\nu_0=\SI{429}{THz}$ is the carrier frequency of the \SI{698}{nm} clock laser.

We simulate the randomized benchmarking experiment by selecting 30 randomly chosen gate strings for each $L=1\dots200$ and utilize Eq.~(\ref{eq:fidelity_fit_decay}) to fit the final $p_e$ of the numerical simulation results to extract the prediction of $F_1^2$ shown in Fig.~\ref{fig:fig4}.
For each gate string we integrate the evolution of the density matrix for 50 different phase traces and average the final $p_e$ to obtain an estimate for the fidelity of the corresponding gate sequence.
Because generating the phase traces for durations of multiple seconds becomes computationally expensive, we re-use the same phase traces for each Clifford gate string. The randomization in selecting the gate strings should provide a sufficient phase-randomization such that the averages extracted by the simulations provide a faithful comparison to our experimental data.

\begin{table*}
\caption{\label{tab:Clifford group} Experimental realization of the elements of single-qubit Clifford group, adapted from Ref.~\cite{Xia2015PRL}. The notation $R_{\Vec{r}}(\theta)$ stands for the rotation operator for rotation along axis $\Vec{r}$ with rotation area $\theta$ and a negative rotation area is realized by a rotation along $-\Vec{r}$ axis. The laser implementation is the operator product $R_3 R_2 R_1$, reading right to left. $U$ is the unitary operator corresponding to each element of the single-qubit Clifford group.}
\begin{ruledtabular}
\begin{tabular}{cccccc}
index & $U$ & $R_3$ & $R_2$ & $R_1$ & $\theta_{total}$ \\ 
\hline
1  & $\begin{pmatrix} 1 & 0 \\ 0 & 1 \end{pmatrix}$ & - & - & - & 0 \\ 
2 & $e^{-i\pi/4}\begin{pmatrix} 1 & 0 \\ 0 & i \end{pmatrix}$ & $R_x(\pi/2)$ & $R_y(\pi/2)$ & $R_x(-\pi/2)$ & $3\pi/2$ \\ 
3 & $-i\begin{pmatrix} 1 & 0 \\ 0 & -1 \end{pmatrix}$ & - & $R_x(\pi)$ & $R_y(\pi)$ & $2\pi$ \\  
4 & $e^{i\pi/4}\begin{pmatrix} 1 & 0 \\ 0 & -i \end{pmatrix}$ & $R_x(\pi/2)$ & $R_y(-\pi/2)$ & $R_x(-\pi/2)$ & $3\pi/2$ \\  
5 & $-1\begin{pmatrix} 0 & 1 \\ -1 & 0 \end{pmatrix}$ & - & - & $R_y(\pi)$ & $\pi$ \\  
6 & $-e^{i\pi/4}\begin{pmatrix} 0 & 1 \\ i & 0 \end{pmatrix}$ & $R_x(\pi/2)$ & $R_y(\pi/2)$ & $R_x(\pi/2)$ & $3\pi/2$ \\  
7 & $-i\begin{pmatrix} 0 & 1 \\ 1 & 0 \end{pmatrix}$ & - & - & $R_x(\pi)$ & $\pi$ \\  
8 & $e^{-i\pi/4}\begin{pmatrix} 0 & 1 \\ -i & 0 \end{pmatrix}$ & $R_x(\pi/2)$ & $R_y(-\pi/2)$ & $R_x(\pi/2)$ & $3\pi/2$ \\  
9 & $\frac{-i}{\sqrt{2}}\begin{pmatrix} 1 & 1 \\ 1 & -1 \end{pmatrix}$ & - & $R_x(\pi)$ & $R_y(\pi/2)$ & $3\pi/2$ \\  
10 & $\frac{1}{\sqrt{2}}\begin{pmatrix} 1 & 1 \\ -1 & 1 \end{pmatrix}$ & - & - & $R_y(-\pi/2)$ & $\pi/2$ \\ 
11 & $\frac{e^{-i\pi/4}}{\sqrt{2}}\begin{pmatrix} 1 & 1 \\ -i & i \end{pmatrix}$ & - & $R_y(-\pi/2)$ & $R_x(\pi/2)$ & $\pi$ \\  
12 & $-\frac{e^{i\pi/4}}{\sqrt{2}}\begin{pmatrix} 1 & 1 \\ i & -i \end{pmatrix}$ & $R_y(-\pi/2)$ & $R_x(\pi/2)$ & $R_x(\pi)$ & $2\pi$ \\  
13 & $\frac{i}{\sqrt{2}}\begin{pmatrix} 1 & -1 \\ -1 & -1 \end{pmatrix}$ & - & $R_x(\pi)$ & $R_y(-\pi/2)$ & $3\pi/2$ \\  
14 & $\frac{e^{-i\pi/4}}{\sqrt{2}}\begin{pmatrix} 1 & -1 \\ i & i \end{pmatrix}$ & - & $R_y(\pi/2)$ & $R_x(-\pi/2)$ & $\pi$ \\  
15 & $\frac{1}{\sqrt{2}}\begin{pmatrix} 1 & -1 \\ 1 & 1 \end{pmatrix}$ & - & - & $R_y(\pi/2)$ & $\pi/2$ \\  
16 & $\frac{e^{i\pi/4}}{\sqrt{2}}\begin{pmatrix} 1 & -1 \\ -i & -i \end{pmatrix}$ & - & $R_y(\pi/2)$ & $R_x(\pi/2)$ & $\pi$ \\  
17 & $\frac{e^{-i\pi/4}}{\sqrt{2}}\begin{pmatrix} 1 & i \\ -1 & i \end{pmatrix}$ & - & $R_x(-\pi/2)$ & $R_y(-\pi/2)$ & $\pi$ \\  
18 & $\frac{e^{i\pi/4}}{\sqrt{2}}\begin{pmatrix} 1 & i \\ 1 & -i \end{pmatrix}$ & - & $R_x(-\pi/2)$ & $R_y(\pi/2)$ & $\pi$ \\  
19 & $\frac{i}{\sqrt{2}}\begin{pmatrix} 1 & i \\ -i & -1 \end{pmatrix}$ & - & $R_x(-\pi/2)$ & $R_y(\pi)$ & $3\pi/2$ \\  
20 & $\frac{1}{\sqrt{2}}\begin{pmatrix} 1 & i \\ i & 1 \end{pmatrix}$ & - & - & $R_x(-\pi/2)$ & $\pi/2$ \\  
21 & $\frac{e^{i\pi/4}}{\sqrt{2}}\begin{pmatrix} 1 & -i \\ -1 & -i \end{pmatrix}$ & - & $R_x(\pi/2)$ & $R_y(-\pi/2)$ & $\pi$ \\  
22 & $\frac{1}{\sqrt{2}}\begin{pmatrix} 1 & -i \\ -i & 1 \end{pmatrix}$ & - & - & $R_x(\pi/2)$ & $\pi/2$ \\  
23 & $\frac{-i}{\sqrt{2}}\begin{pmatrix} 1 & -i \\ i & -1 \end{pmatrix}$ & - & $R_x(\pi/2)$ & $R_y(\pi)$ & $3\pi/2$ \\  
24 & $\frac{e^{-i\pi/4}}{\sqrt{2}}\begin{pmatrix} 1 & -i \\ 1 & i \end{pmatrix}$ & - & $R_x(\pi/2)$ & $R_y(\pi/2)$ & $\pi$ \\  
\end{tabular}
\end{ruledtabular}
\end{table*}

\clearpage
\bibliography{main}% Produces the bibliography via BibTeX.

\end{document}